\documentclass[11pt]{article}
\usepackage{moriond,epsfig,cite}

\def\be{\begin{equation}}
\def\ee{\end{equation}}
\def\bea{\begin{eqnarray}}
\def\eea{\end{eqnarray}}

\begin{document}

\begin{flushright}
CERN-PH-TH/2008-147\\
SLAC-PUB-13293\\
FERMILAB-CONF-08-228-T\\
SFB/CPP-08-46\\
PITHA 08/15
\end{flushright}

\vspace*{1cm}
\title{THE ROLE OF COLLINEAR PHOTONS IN THE RARE DECAY $\bar{B} \to X_s \ell^+ \ell^-$}

\author{T. HUBER$^{1}$, T. HURTH$^{2,3}$, E. LUNGHI$^{4}$}

\address{$^1$~Institut f\"ur Theoretische Physik E, RWTH Aachen University,\\
D - 52056 Aachen,Germany\\
$^2$~CERN, Dept. of Physics, Theory Division, CH-1211 Geneva, Switzerland\\
$^3$~SLAC, Stanford University, Stanford, CA 94309, USA\\ 
$^4$~Fermi National Accelerator Laboratory, P.O.Box 500, Batavia, IL 60510, U.S.A.}

\maketitle\abstracts{We review the phenomenology of the rare decay $\bar{B} \to X_s \ell^+ \ell^-$. We present the results of a detailed phenomenological analysis and discuss the r\^ole of the decay in the search for new physics at
present and future colliders. Moreover, we extensively elaborate on the size of electromagnetic logarithms $\ln(m_b^2/m_{\ell}^2)$ in view of experimental cuts. We point out the differences in the analyses of BaBar and Belle and give
suggestions on how to treat collinear photons in the experimental analyses. These recommendations correspond precisely to theoretical prescriptions and can be combined with measurements performed at a Super-$B$ factory.}

\section{Introduction}
In recent years, several flavour facilities as well as the experiments at the Fermilab Tevatron have accumulated a large set of data and have confirmed the CKM mechanism~\cite{Cabibbo:1963yz,Kobayashi:1973fv} of quark flavour mixing and CP violation with tremendous success. Besides the determination of the parameters of the CKM matrix and the unitarity triangle, one major goal has been the search for new phyics (NP) beyond the Standard Model (SM). Among the prime candidates for this search are observables related to rare, flavour-changing neutral current (FCNC) decays of $B$, $D$, and $K$ mesons. These decays probe the SM directly at the loop level and are therefore, via virtual effects, sensitive to scales presently not accessible at direct collider experiments. With the start of the LHC being within eyespot, the search for new physics via direct production of new degrees of freedom will also become feasible in the near future, and the situation calls for an interplay between flavour and collider physics~\cite{delAguila:2008iz,Buchalla:2008jp,Raidal:2008jk}.

\begin{figure}[t]
\hspace*{18pt}\scalebox{0.58}{\includegraphics[scale=.80]{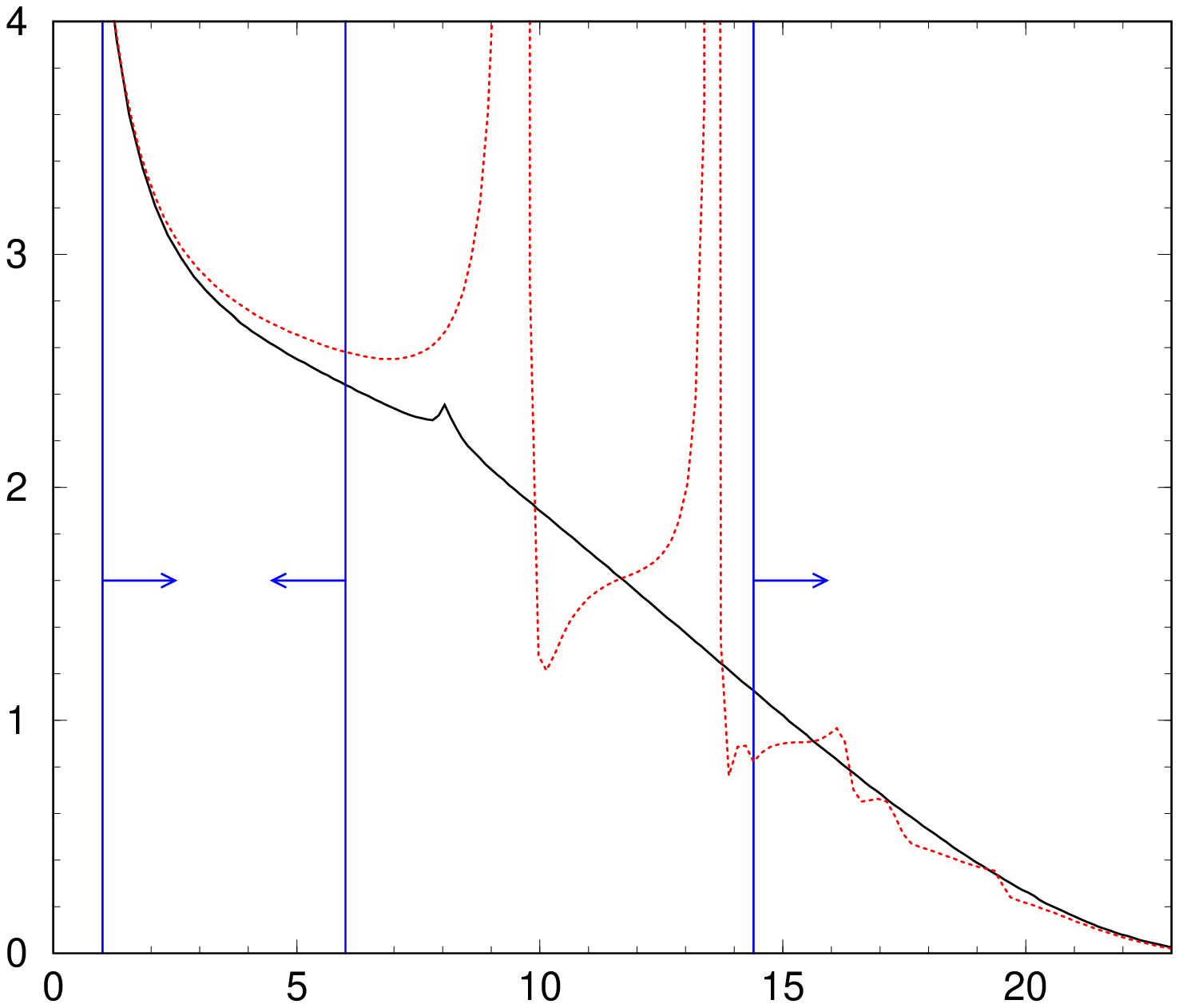}}
\begin{picture}(0,0)%
\setlength{\unitlength}{1pt}%
\put(-67,80){{\raisebox{-15mm}[0mm][0mm]{%
           \makebox[-100mm]{\scalebox{0.8}{\rotatebox{90}{$\frac{d \rm{BR}_{\ell\ell}}{d
q^2} \times 10^7~ \left [ {\rm GeV}^{-2} \right ]$}}}}}}%
\put(-40,0){{\raisebox{-3.5mm}[0mm][0mm]{%
           \makebox[-53mm]{\scalebox{0.7}{$q^2~ \left [ {\rm GeV}^2 \right
]$}}}}}%  
\end{picture}%
\hspace*{18pt}\scalebox{0.58}{\includegraphics[scale=.80]{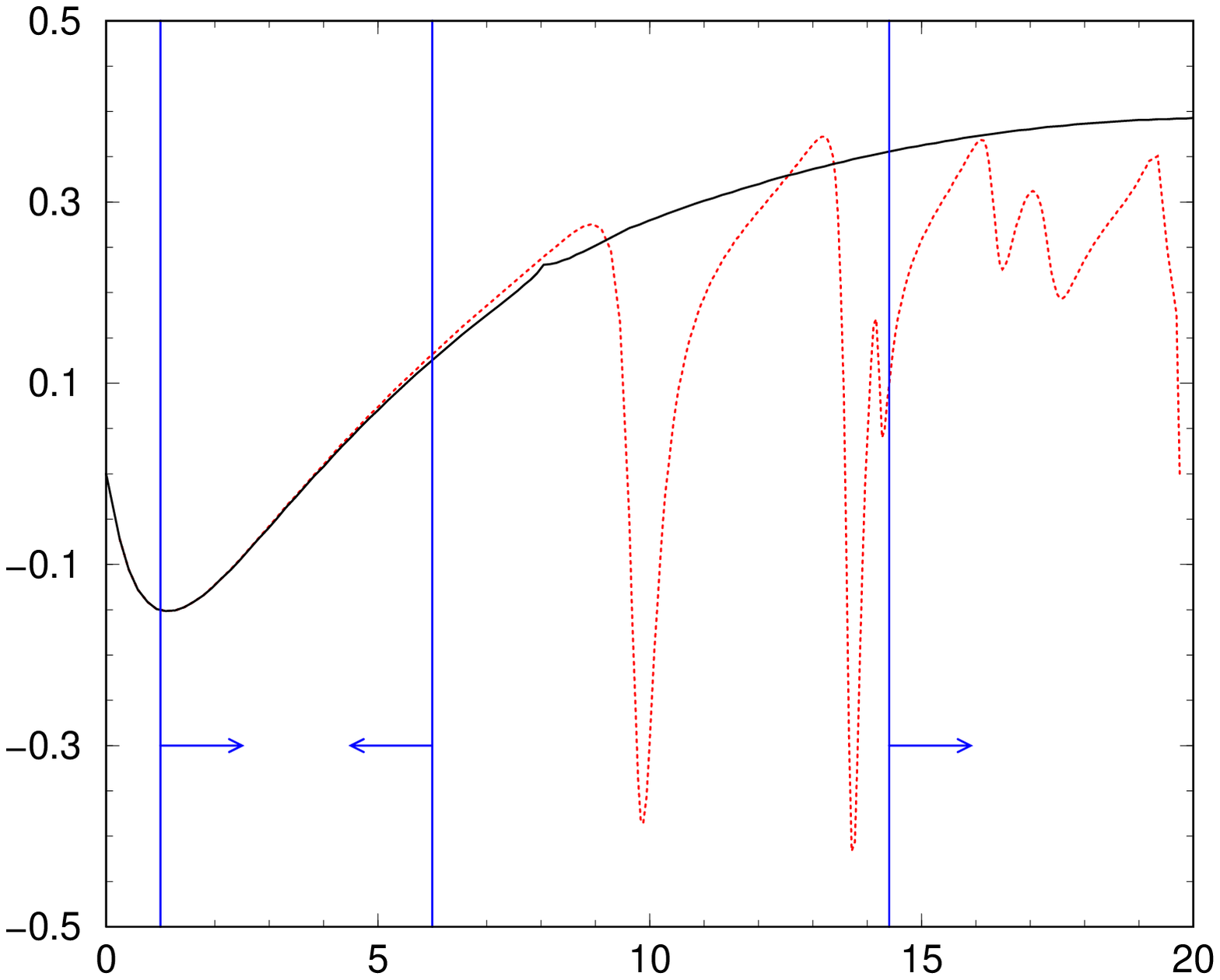}}
\begin{picture}(0,0)%
\setlength{\unitlength}{1pt}%
\put(-76,73){{\raisebox{0mm}[0mm][0mm]{%
           \makebox[-100mm]{\scalebox{0.8}{\rotatebox{90}{$A_{FB}(q^2)$}}}}}}%
\put(-30,0){{\raisebox{-3.5mm}[0mm][0mm]{%
           \makebox[-53mm]{\scalebox{0.7}{$q^2~ \left [ {\rm GeV}^2 \right]$}}}}}%  
\end{picture}%
\vskip3mm
\caption{Left panel: Differential branching ratio (BR) as a function of $q^2$ without (solid, black) and with (dotted, red) factorizable $c\bar c$ corrections from the KS approach. The vertical (blue) lines with the arrows indicate the perturbative windows. Right panel: Forward-backward asymmetry (FBA) as a function of $q^2$. The meaning of the lines is the same as before~\cite{Adrian1}.\label{fig:BRandFBA}}
\end{figure}

Among inclusive flavour-changing neutral current (FCNC) processes (for a review see~\cite{Hurth:2007xa,Hurth:2003vb}), the inclusive $\bar B \rightarrow X_s \ell^+\ell^-$  decay represents an important test  of the SM, complementary to the inclusive $\bar B \rightarrow X_s \gamma$ decay.  The two most attractive kinematic observables are the dilepton invariant mass spectrum and the forward-backward asymmetry (FBA), see Fig.~\ref{fig:BRandFBA}. In the so-called `perturbative $q^2$-windows', namely in the low-dilepton-mass  region $1\,{\rm GeV}^2 < q^2 = m_{\ell\ell}^2   < 6\,{\rm GeV}^2$, and also in the high-dilepton-mass region with $q^2 > 14.4\,{\rm GeV}^2$, theoretical predictions for the invariant mass spectrum are dominated by the perturbative contributions, and a theoretical precision of order $10\%$ is in principle possible.
\section{Theoretical framework and Phenomenological results}
The computation of observables of rare decays in flavour physics involves two widely separated scales $M_H \gg M_L$, with $M_H \simeq {\cal O} (M_W,M_Z,m_t)$ and $M_L \simeq {\cal O}(m_b)$, entailing the need for resummation of the occurring large logarithms $\ln(M_H^2/M_L^2)$. This is most conveniently done in the framework of an effective theory where the top-quark as well as the heavy electroweak gauge bosons are integrated out. In this framework the occurring large logarithms can be resummed order by order in $\alpha_s$ by means of techniques of the renormalization group-improved perturbation theory. The relevant Lagrangian density can be found, e.g., in Ref.~\cite{Huber:2005ig}.

The calculations in $\bar B \to X_s \ell^+ \ell^-$ have achieved a very sophisticated level. The recently calculated NNLL QCD contributions~\cite{MISIAKBOBETH,Asa1,Asatryan:2002iy,Adrian2,Asatrian:2002va,Ghinculov:2003bx,Adrian1,Gambinonew,Asatrian:2003yk,Gorbahn:2004my} have significantly improved the sensitivity of the inclusive $\bar B \rightarrow X_s \ell^+ \ell^-$ decay in  testing extensions of the SM in the sector of flavour dynamics. In particular, the value of the dilepton invariant mass $q^2_0$ for which the differential FBA vanishes is one of the most precise predictions in flavour physics with a theoretical uncertainty of order $5\%$. This well corresponds to the expected experimental sensitivity of $4-6 \%$ at the proposed Super-$B$ factories~\cite{Akeroyd:2004mj,Hewett:2004tv,Bona:2007qt,Browder:2007gg}. Also non-perturbative corrections scaling with $1/m_b^2$, $1/m_b^3$, or $1/m_c^2$\,\,\cite{Falk,Alineu,Savagenew,Buchalla:1997ky,buchallanewnew,Bauer,Ligeti:2007sn} have to be  taken into account. Moreover, factorizable long-distance contributions away from the resonance peaks are important; here using the Kr\"uger-Sehgal approach~\cite{KS} avoids the  problem of double-counting.

In the high-$q^2$  region, one encounters the breakdown of the heavy-mass expansion at the endpoint; while the partonic contribution vanishes in the end-point, the $1/m_b^2$ and $1/m_b^3$ corrections tend towards a non-zero value. However, for an integrated high-$q^2$ spectrum an effective expansion is found in inverse powers of $m_b^{\rm eff} = m_b \times (1 - \sqrt{\hat s_{\rm min}})$ rather than $m_b$~\cite{Neubert,BLK}.

\begin{figure}[t]
\begin{center}
\includegraphics[scale=.76]{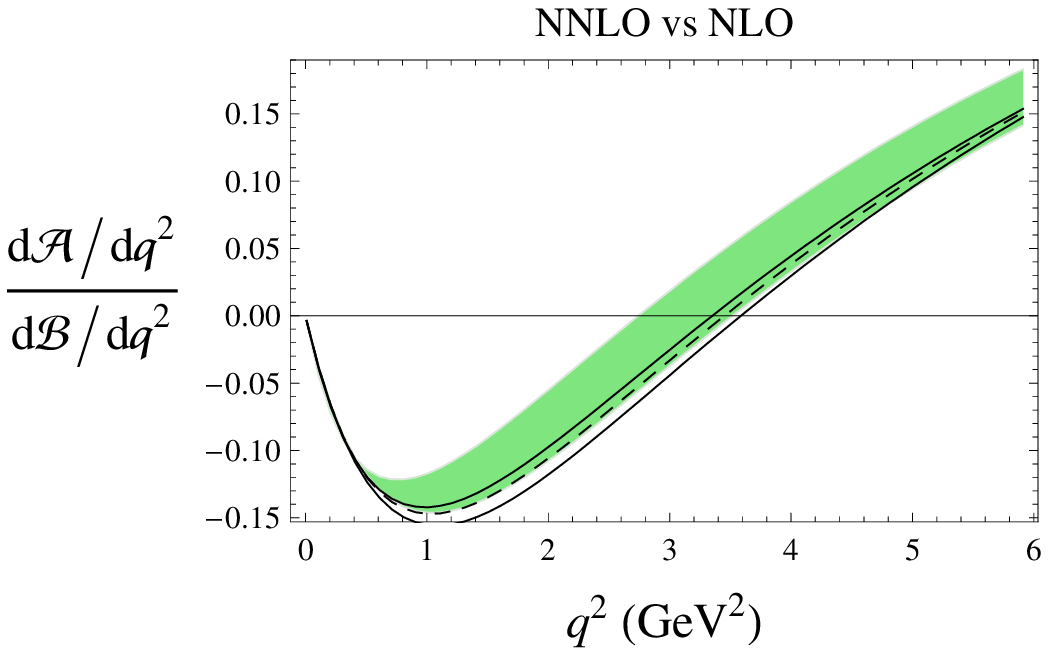}
\includegraphics[scale=.76]{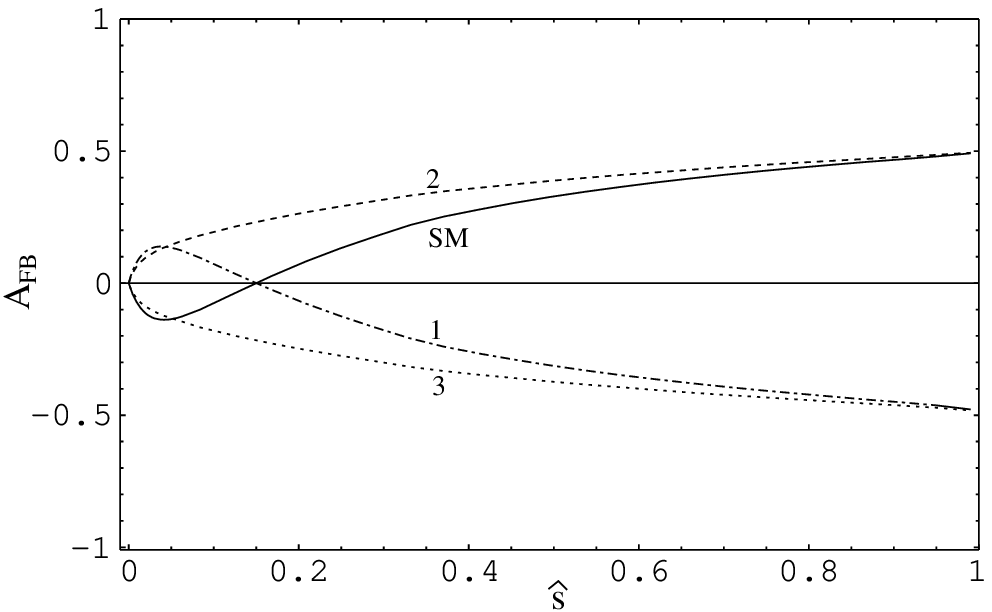}
\end{center}
\caption{Left panel: $\mu_b$-dependence of the forward backward asymmetry for the muonic final state. The lines are the NNLO QCD result; the dashed line corresponds to $\mu_b = 5$~GeV, and the solid lines to $\mu_b = 2.5,10$~GeV. The shaded area is the region spanned by the NLO asymmetry for $2.5$~GeV $<\mu_b< 10$ GeV. Right panel: Solid: FBA as a function of the lepton invariant mass. Curve 2: Reversed sign of $C_7$ w.r.t. SM. Curves 1,3: Sign of $C_{10}$ reversed in addition to curves SM,2 respectively~\cite{Ali:2002jg}.}
\label{fig:AFBlowsplot}
\end{figure}

Recently, further refinements were presented such as the NLO QED two-loop corrections to the Wilson coefficients whose size is of order $2\%$~\cite{Gambinonew}. Furthermore, it was shown that in the QED one-loop corrections to matrix elements large collinear logarithms of the form $\log(m_b^2/m^2_{\rm lepton})$ survive integration if only a restricted part of the dilepton mass spectrum is considered. This adds another $+2\%$ contribution in the low-$q^2$ region for ${\cal B} (\bar B\to X_s \mu^+\mu^-)$ and results in~\cite{Huber:2005ig}
\be\label{muonBR} 
{\cal B} (\bar B\to X_s \ell^+\ell^-)_{[1 < q^2/{\rm GeV}^2 < 6]}  = \cases{
(  1.59  \pm 0.11 ) \times 10^{-6} & $\ell=\mu$ \cr
(  1.64  \pm 0.11 ) \times 10^{-6} & $\ell=e \, .$ \cr}
\ee
We will elaborate more on the difference between electron and muon channel in section~\ref{sec:collineargamma}. In Ref.~\cite{Huber:2007vv} also the results for the high-$q^2$ region and for the FBA were derived. The result for the branching ratio (BR) in the high-$q^2$ region reads
\be\label{muonBRhighs} 
{\cal B} (\bar B\to X_s \ell^+\ell^-)_{[q^2> 14.4 \, {\rm GeV}^2]}  = \cases{
2.40 \times 10^{-7} \; (1^{+0.29}_{-0.26} ) & $\ell=\mu$ \cr
2.09 \times 10^{-7} \; (1^{+0.32}_{-0.30} ) & $\ell=e \, .$ \cr}
\ee
In this case the relative impact of the collinear QED logarithm is about $-8$\% ($-20$\%) for muons (electrons) and therefore much larger than in the low-$q^2$ region due to the steep decrease of the differential decay width at large $q^2$. The large error in Eq.~(\ref{muonBRhighs}) is mainly due to the sizable uncertainties in the parameters that enter the $O(1/m_b^3)$ non-perturbative corrections. As was pointed out in Ref.~\cite{Ligeti:2007sn} the error can be
significantly decreased by normalizing the $\bar B \rightarrow X_s \ell^+ \ell^-$ decay rate to the semileptonic $\bar B \rightarrow X_u \ell\bar\nu$ decay rate {\textit{with the same $q^2$ cut}}. For a lower cut of $q_0^2 =
14.4$~GeV$^2$ this leads to~\cite{Huber:2007vv}
\be\label{muonR} 
{\cal R}^{\ell\ell}(\hat s_0) = 
\int_{\hat s_0}^1 {\rm d} \hat s \, {{\rm d} {\Gamma} (\bar B\to X_s \ell^+\ell^-) \over {\rm d} \hat s} \bigg/
\int_{\hat s_0}^1 {\rm d} \hat s \, {{\rm d} {\Gamma} (\bar B^0\to X_u \ell \nu) \over {\rm d} \hat s} 
 = \cases{
2.29 \times 10^{-3} ( 1 \pm 0.13) & $\ell=\mu$ \cr
1.94 \times 10^{-3} ( 1 \pm 0.16) & $\ell=e \, .$ \cr}
\ee
where $\hat s = q^2/m_b^2$. The uncertainties from poorly known $O(1/m_b^3)$ power corrections are now under control; the largest source of error is $V_{ub}$. The zero of the FBA is found to be at
\be\label{eq:muonzero}
(q_0^2)_{\ell\ell} = \cases{
( 3.50 \pm 0.12) \, {\rm GeV}^2  &  $\ell=\mu$ \cr
( 3.38 \pm 0.11) \, {\rm GeV}^2  &  $\ell=e\, .$ \cr}
\ee
The error is about 3\% but includes parametric and perturbative uncertainties only. However, unknown subleading non-perturbative corrections of order $O(\alpha_s \Lambda/m_b)$, which are estimated to give an additional uncertainty of order 5\%, have to be added in addition. It is often argued that especially the small $\mu$ dependence at the zero is an accident and should  be increased by hand. However, by comparing the NLO-QCD with the NNLO-QCD result one can clearly see that the $\mu$ dependence is a reasonable reflection of the perturbative error, see left panel in Fig.~\ref{fig:AFBlowsplot}. Moreover, the zero is stable under change of the $b$ quark mass scheme; the variation is below $2$\% when switching from 1S to $\overline{\rm MS}$ or pole scheme.
\section{New physics sensitivities}
By the end of the $B$ factories the fully differential shape of the branching
ratio and FBA will not be accessible, contrary to their integrals over bins in
the low-$q^2$ region, which are usually chosen to be $q^2 \in [1,3.5]$~GeV$^2$
and $q^2 \in [3.5,6]$~GeV$^2$~\cite{Lee:2006gs}. These quantities will already
allow to discriminate between different NP scenarios, see right panel in
Fig.~\ref{fig:AFBlowsplot} as well as
Refs.~\cite{Akeroyd:2004mj,Ali:2002jg,Gambino:2004mv}. In the SM the integrated
FBA over various bins in the low-$q^2$ region reads~\cite{Huber:2007vv}
\be
\label{integratedAFB}
\bar{\cal A}_{\ell\ell}^{(1,3.5)}  =  \cases{( -9.09 \pm 0.91 )\%  \cr ( -8.14 \pm 0.87 )\% \cr} \; ,
\bar{\cal A}_{\ell\ell}^{(3.5,6)}  =  \cases{ ( 7.80 \pm 0.76 )\%  \cr ( 8.27 \pm 0.69 )\% \cr } \; ,
\bar{\cal A}_{\ell\ell}^{(1,6)}  =  \cases{  ( -1.50 \pm 0.90 )\%  \cr ( -0.86 \pm 0.85 )\% \, ,\cr }
\ee
where the upper~(lower) line corresponds to the muon~(electron) final state. The relative errors in the respective bins are considerably smaller than for the entire low-$q^2$ region since the respective values in each bin are similar in
size and of opposite sign. The first two numbers do not add up to the third one for normalization reasons~\cite{Huber:2007vv}. In their analysis the authors of Ref.~\cite{Lee:2006gs} consider the three linearly independent quantities
that can be extracted from the double differential decay width ($z =\cos\theta$)~\footnote{$\theta$ is the angle between the positively charged lepton and the $\bar B$ in the c.m.s. of the lepton pair.}
\be
d^2\Gamma/(dq^2 dz) = 3/8 \, \left[(1+z^2) \, H_T(q^2) \, + \, 2 \, z \, H_A(q^2) +2 \, (1-z^2) \, H_L(q^2)\right] \; ,
\ee
where
\be
d\Gamma/dq^2 = H_T(q^2) \, +H_L(q^2)\, , \qquad dA_{\rm{FB}}/dq^2 = 3/4 \, H_A(q^2) \, ,
\ee
integrated over the aforementioned bins in the low-$q^2$ region. They are able to put contraints on the Wilson Coefficients $C_9$ and $C_{10}$ by imposing a negative $C_7$ whose magnitude is taken from ${\cal B}(\bar B \to X_s \gamma)$. Therefore, if the statistics allows, we highly encourage the experimental groups to present their results separately for the three linearly independent observables and for the two bins in the low-$q^2$ region. Hence the measurements of the branching ratio and the FBA, in addition to the $\bar B \rightarrow X_s \gamma$ branching ratio, will allow to fix magnitude and sign of all relevant Wilson coefficients in the SM and to put constraints on the parameter space of NP models.

New physics might also affect the high-scale Wilson Coefficients in such a way that they aquire additional phases. In Refs.~\cite{Huber:2005ig,Huber:2007vv} we give the results for the branching ratio and FBA in terms of generic high-scale Wilson Coefficients. These results may serve to constrain the parameter space of NP models and are also of interest in other processes~\cite{Hou}.

\section{Collinear photons}\label{sec:collineargamma}
After including the NLO QED matrix elements, the electron and muon channels receive different contributions due to terms involving $\ln(m_b^2/m_\ell^2)$. This is the only source of the difference between these two channels. We
emphasize that the results we present in Eqs.~(\ref{muonBR})--(\ref{integratedAFB}) correspond to the process $\bar B \to X_s \ell^+ \ell^-$ in which QED photons are included in the $X_s$ system and the di-lepton invariant mass does not
contain any photon, i.e. $q^2 = (p_{\ell^+} + p_{\ell^-})^2$. This would be exactly the case in a fully inclusive analysis using the recoil technique (such an analysis would require a Super-$B$
machine~\cite{Akeroyd:2004mj,Hewett:2004tv,Bona:2007qt,Browder:2007gg}).

However, as already pointed out in Refs.~\cite{Huber:2005ig,Huber:2007vv}, the presence of the logarithm is strictly related to the definition of the dilepton invariant mass. If {\textit{all}} photons emitted by the final state on-shell
leptons are included in the definition of $q^2$: $(p_{\ell^+}+p_{\ell^-})^2\rightarrow (p_{\ell^+}+p_{\ell^-}+p_\gamma)^2$ then the electromagnetic logarithm $\ln(m_b^2/m_\ell^2)$ would be absent due to the
Lee--Kinoshita--Nauenberg theorem~\cite{Lee:1964is,Kinoshita:1962ur}, and hence its effect would
disappear.\footnote{We note here that if the photons emitted by the final $X_s$ system were also included in the $q^2$ definition then there would be an additional $\ln(m_s^2/m_b^2)$.} Only if all these photons are included in the $X_s$
system and not in the $q^2$ (i.e.\ if a perfect separation of leptons and collinear photons is achieved) our expressions containing $\ln(m_b^2/m_\ell^2)$ are directly applicable. We emphasize that by collinear photons we denote only
photons that are emitted in the vacuum. No photons emitted in the external magnetic field of the detector nor from interaction with the detector material are considered.

In the BaBar and Belle experiments the inclusive decay is measured as a sum over exclusive states. Moreover, the treatment of the QED radiation is different~\cite{privatebabar,privatebelle}. As  a consequence the log-enhanced QED
corrections are not directly applicable to the present experimental results and have to be modified. Let us elaborate on this.
 
For what concerns the di-muon final state, both BaBar and Belle do not include hard collinear photons in the $q^2$ definition nor in the final $X_s$ system. So the collinear logarithms associated to real photon
emission are absent. Hence, the theoretical prediction should contain only the logarithms appearing in the calculation of virtual photon exchanges plus real unresolved (i.e.\ soft) emission. In
Eqs.~(\ref{muonBR})--(\ref{integratedAFB}) we added all real and virtual effects. Because of the details of the regularization procedure we used in Refs.~\cite{Huber:2005ig,Huber:2007vv} (dimensional regularization
for both soft and collinear singularities), a dedicated calculation is required in order to disentangle soft and collinear contributions.

For what concerns the di-electron final state, both experiments include photons in a cone whose opening angle $\theta$ is about $(35-50)
\; {\rm mrad}$ around the lepton directions. The current BaBar exclusive, and projected inclusive, $s \ell^+ \ell^-$ analyses
use only the single most energetic photon above the minimum energy threshold (see below) lying in the cone around the electron.
For photons emitted in this cone we have $m_\ell^2\leq(p_\ell + p_\gamma)^2 \leq
\Lambda^2 \simeq 2 E_\ell^2 (1-\cos\theta)$, where $E_\ell$ is the energy of the lepton, usually of order $m_b/2$. Using $E_\ell =
m_b/2$ and $\theta=45$~mrad, $\Lambda$ is found to be of order $m_\mu$.  Thus, as discussed in
Refs.~\cite{Huber:2005ig,Huber:2007vv}, if the photons within the cone were included in the $q^2$ definition and the collinear photons
outside the cone in the final $X_s$ system, our results would be directly applicable with $\ln(m_b^2/m_e^2)$ replaced by
$\ln(m_b^2/\Lambda^2)$ where $\Lambda \sim O(m_\mu)$.

However, the experimental situation is even more complicated. At BaBar and at Belle the collinear photons outside the cone are not
included at all neither in the $q^2$ nor in the $X_s$ contribution. At Belle events with energetic photons
($>100$~MeV) or less energetic photons ($20$--$100$~MeV) are effectively vetoed or suppressed, respectively, by requiring energies and
the momenta of the final state particles (two leptons, a $K^{(*)}$ and up to four pions) to add up to the B meson energy and momentum.
At BaBar energetic photons outside the cone are effectively vetoed by requiring the momenta of the final state particles (two leptons, a
$K^{(*)}$ and up to three pions) to add up to the $B$ meson momentum. Moreover, the two experiments differ on the definition of the
di-electron invariant mass squared, $q^2$: BaBar includes the collinear photons inside the cone in the $q^2$, while Belle does not. The
first condition requires the subtraction of the BR for $\bar B\to X_s e^+ e^- \gamma$ integrated over photons lying outside of the cone:
this effect is proportional to $\log(m_b^2/\Lambda^2)$ and can be calculated only by means of a numerical integration over the phase
space. The second one implies that the $\log(m_b^2/m_e^2)$ is absent in the BaBar case and present in the Belle one.

Finally we remind that photons which are resolved but whose energy is below an experiment--dependent threshold of about $30\; (20)$~MeV in the lab frame at BaBar (Belle), are not observed, i.e., are not included in
the di-lepton nor in the $X_s$ system, and effectively shift the energy of the reconstructed $B$  mesons. However, most likely these events are still considered to be a candidate inside the selection window. These events
therefore introduce on the measured branching ratio a logarithmic dependence on the aforementioned energy threshold. This dependence can be computed by means of the soft photon approximation.
In this way all the diagrams factorize into tree--level times photon emission. Also the phase space factorizes, and the integral will just give the soft and soft-collinear singularities. This kernel can be integrated
up to the soft cut and when added to the virtual terms, the soft and soft-collinear singularities will drop out and will be replaced by the logarithm of the cut.

The conclusion is that the theoretical predictions for the measurements published by both experiments are affected by large collinear logarithms whose precise size has to be determined by separate (possibly numerical) calculations which
take into account the experimental cuts on the photon energy and the opening angle of the collinear cone, as well as the boost from the lab frame into the rest frame of the $\bar B$. We urge both collaborations to converge on a unique
definition of these processes. We recommend to search and include {\em all photons} in the final state and to define the di-lepton invariant mass {\em without} the photon momentum (i.e. $q^2 = (p_{\ell^+}+ p_{\ell^-})^2$): in this way
the measured rate and asymmetries would correspond precisely to our theoretical predictions and could be combined with future measurements performed with the recoil technique. Alternatively, the exclusion of all collinear photons from
the measurement of the rate would allow a fully analytical calculation of the theoretical prediction (except for of the soft photon cut dependence).

%{\bf Kevin Flood made a remark today here at CERN that including the collinear photons in the analysis (namely in the $X_s$ system) in the present analsyis would enlarge the experimental error quite a bit.?} 

\section*{Acknowledgments}
We would like to thank the organizers of Moriond QCD 2008 for creating a pleasant and inspiring atmosphere. This work was supported by DFG, SFB/TR 9.
Fermilab is operated by Fermi Research Alliance, LLC under Contract No. DE-AC02-07CH11359 with the United States Department of Energy. We are grateful to the experimental groups of BaBar and Belle for discussions on the treatment of
collinear photons. Discussions with P.~Zerwas are also gratefully acknowledged. T.~Huber acknowledges receipt of a grant from the EU ``Marie Curie'' Programme.

\end{document}